\documentclass[
aps,
preprint,
superscriptaddress,
showpacs,
showkeys,
longbibliography
]{revtex4-2}

\usepackage[utf8]{inputenc}
\usepackage[T1]{fontenc}
\usepackage{amsmath,amssymb,amsfonts}
\usepackage{graphicx}
\usepackage{multirow}
\usepackage{kotex}
\usepackage{placeins}

\renewcommand{\arraystretch}{1.5}

\begin{document}
	
	\title{Spatially explicit population modeling with habitat-preferential movement of \textit{Glis glis} (edible dormouse)}
	
	\author{Jeehye Choi}
	\affiliation{Advanced-Basic-Convergence Research Institute, Chungbuk National University, Cheongju, Chungbuk 28644, Korea}
	
	\author{KyoungEun Lee}
	\email{kelee25@gmail.com}
	\affiliation{National Institute of Ecology (NIE), Seocheon, Chungnam 33657, Korea}
	\affiliation{Korea Environment Institute (KEI), Sejong 30147, Korea}
	
	\author{Do-Hun Lee}
	\affiliation{National Institute of Ecology (NIE), Seocheon, Chungnam 33657, Korea}
	
	\author{Byungjoon Min}
	\affiliation{Advanced-Basic-Convergence Research Institute, Chungbuk National University, Cheongju, Chungbuk 28644, Korea}
	\affiliation{Department of Physics, Chungbuk National University, Cheongju, Chungbuk 28644, Korea}
	
	\author{Tae-Soo Chon}
	\email{tschon.chon@gmail.com}
	\affiliation{Office of Research and University-industry Cooperation, Kyung Hee University, Seoul 02447, Korea}
	\affiliation{Research Institute of Computer, Information and Communication, Pusan National University, Busan 46244, Korea}
	
	\date[]{}

	\begin{abstract}
		
		We develop a spatially explicit dynamical model for the dispersal of the edible dormouse, \textit{Glis glis}, in a heterogeneous landscape. Population growth occurs within lattice cells, while movement between neighboring cells is governed by logistic functions representing density-dependent emigration and habitat preference. The model includes growth rate, carrying capacity, Allee threshold, home-range scale, and land-cover-based habitat suitability. Simulations over 100 years show that greater sensitivity to habitat preference tends to produce localized distributions with higher mean population density, whereas greater sensitivity to density dependence tends to yield larger occupied areas with lower mean population density. The model links local movement rules to emergent population-level patterns and offers a transparent, process-based framework for exploring potential invasive-species dispersal under field conditions where occurrence data are limited.
		
	\end{abstract}
	
\pacs{87.23.Cc, 05.10.-a}

\keywords{spatially explicit model, population dispersal, density-dependent movement, habitat preference, invasive species}
	
	\maketitle

\section{Introduction}\label{sec:intro}
Population dynamics are a key component of ecological studies, focusing on both theoretical aspects like ecological stability  \citep{Begon1990ecology} and practical issues such as managing invasive species \citep{Veitch2002turning,Arim2006spread}. 
Using population models, researchers can gain insights into the mechanisms that drive changes in species densities and distributions while tackling practical challenges.
These models serve as vital tools for predicting the spread of invasive species and supporting the conservation of at-risk species by offering reliable forecasts \citep{Eiswerth2002managing,Gallien2010predicting,Lohr2017modeling,Lustig2017modeling}.

Population modeling in ecology has evolved over time, incorporating diverse methods and ideas. 
Among the earliest models is the logistic growth model \citep{Verhulst1845}, which describes population growth and eventual saturation but lacks consideration of dispersal. 
To address this, researchers expanded the model to include dispersal phases like invasion, establishment and proliferation in space \citep{Shigesada2002invasion}, using approaches such as reaction-diffusion processes \citep{Cosner2008}. 
While early studies assumed random movement \citep{Skellam1951random,Hastings2005spatial}, real-world animal movement is shaped by environmental factors and species-specific behaviors \citep{Mueller2008search}.

Since species movement is significantly influenced by site-specific properties under real conditions, it is imperative to include habitat preferences within heterogeneous environments \citep{Geuse1985distribution,Bowne1999effects}. 
To address this, spatially explicit models have been developed to integrate population dynamics and environmental heterogeneity within a spatiotemporal framework \citep{Dunning1995Spatially,book_Tilman1997,book_Malchow2007,Jeschke2012Support,Deangelis2017spatially}. 
Furthermore, animal movement patterns are influenced not only by spatial factors but also by the density of animals in a region \citep{Murray1967Dispersal,Waser1985Does,Matthysen2005density}. 
For instance, higher densities can promote spatial movement of individuals due to increased competition for resources. 
However, few spatially explicit models have concurrently considered density-dependent movement and habitat preference.

In this study, we propose a spatially explicit model that captures dispersal processes by incorporating the ecological traits of the species. These ecological traits include potential growth rate, home range, carrying capacity, and habitat preference. The landscape is divided into spatial units based on the home range. The number of individuals moving between spatial locations is determined by two components: the number leaving the current cell, which depends on the difference between local population size and carrying capacity, and the destination-selection rate, which depends on differences in habitat preference. Both components are represented using logistic functions.

We applied our model to examine the potential dispersal pathways of the edible dormouse (\textit{G. glis}), which was listed among the 200 alert alien species in Korea \citep{book_nie}. 
Our model demonstrates that spatial dispersal patterns, including the occupied area and long-term population density, are strongly influenced by both habitat preference and density dependence. 

This paper is organized as follows. In Sec.~\ref{sec:model} we introduce our spatially explicit model 
to incorporate both habitat preference and density dependence.
We also provide information on the model species, study area, and specific parameters in the model.
We then present numerical results of our model and examine the effects 
of the parameters for movement in Sec.~\ref{sec:result}. 
Finally, we discuss the implications of our study and provide directions for future research in Sec.~\ref{sec:conclusion}.

\section{Model and Methods}\label{sec:model}

Our focus is on developing a spatially explicit model for representing the dispersal of the edible dormouse population while incorporating local spatial information. Considering the potential invasion of the Korean Peninsula by this species, we propose a spatially explicit model that integrates parameters related to population dispersal mechanisms, including potential growth rate and density dependence. In the following subsections, we first introduce our model for the dispersal of a non-native invasive species. We then provide information on the target invasive species and the study area in which the numerical simulations were conducted. We also describe the method used to assess habitat-preference scores, together with details of the parameter sets used in our model.

\subsection{Spatially Explicit Model}

\begin{figure}[]\centering
\includegraphics[width=0.3\columnwidth]{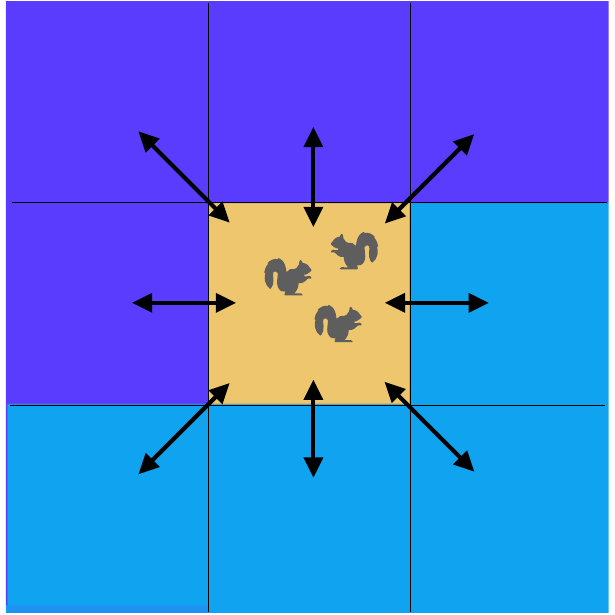}
\caption{Population movement on the two-dimensional lattice}
\end{figure}

Our model consists of two main components: population growth within a cell and population movement between cells. Population growth within a given cell is described by a logistic-type growth function incorporating an Allee effect. For population movement, we incorporate the tendency of individuals to enter a new cell according to habitat preference and to leave the current cell according to density dependence. Population movement in our model is based on two assumptions. First, invasive species tend to migrate toward more suitable habitats, where environmental conditions favor their survival and reproduction. Second, when the local population size approaches or exceeds the carrying capacity---defined as the maximum population level that the environment can sustain---individuals are compelled to disperse to new areas.

The local population size in cell $i$ is denoted by $N_i(t)$. As the model is deterministic,
$N_i(t)$ is treated as a continuous-valued local abundance rather than an integer-valued individual count. In this study, we use a two-dimensional lattice composed of discrete cells to model population dispersal. The population $N_i(t)$ evolves according to the following dynamical equation:

\begin{align}
\frac{\mathit{\Delta}N_i}{\mathit{\Delta}t}&= g_i+ \sum_{j \in \partial i} \left( L_j c_{ji} - L_i c_{ij} \right),
\label{rate_eq}
\end{align}

where $g_i$ is the population-growth function in cell $i$, $\partial i$ denotes the set of neighboring cells of cell $i$, $L_i$ is the number of individuals leaving cell $i$, and $c_{ij}$ is the movement rate from cell $i$ to cell $j$. The first term in the equation represents population growth within each cell, and the second term represents the movement of individuals between cells. The time evolution of $N_i$ at all sites proceeds simultaneously from the initial state. Here, $\partial i$ comprises the eight adjacent cells,
including diagonal neighbors, on the square lattice, as illustrated in Fig.~1. The details of each term in Eq.~\eqref{rate_eq} are explained below.

For population growth in cell $i$, we use a well-known logistic growth equation with an Allee effect \citep{Lewis1993allee,Jung2020Allee}. The population-growth function $g_i$ is given by

\begin{align}
g_i = r N_i \left( \frac{K_i-N_i}{K_i}\right) \left(\frac{N_i -A}{K_i}\right),
\label{growth_eq}
\end{align}

where $r$ is the potential growth rate, $K_i$ represents the local carrying capacity, which
is proportional to the habitat score, and $A$ is the Allee-effect threshold used in the
phenomenological growth function  \citep{Allee1931Animal,Amarasekare1998Allee}. For cells with $K_i>A$, the function $g_i$ describes a strong Allee effect. The local growth term is negative for $0<N_i<A$, positive for $A<N_i<K_i$, and negative for $N_i>K_i$. Thus, $A$ represents a lower threshold for positive local population growth when $K_i>A$.

Because $K_i$ varies according to habitat conditions whereas $A$ is fixed, some cells in low-suitability environments, such as industrial areas and transportation facilities (with raw habitat-preference scores of 1.15 and 1.24, respectively; see Table~\ref{table} in Appendix~A), have $0<K_i<A$. After normalization using Eq.~(5), their corresponding carrying capacities are approximately 0.65 and 1.64 individuals per cell, respectively, both below $A=2$.

The same growth function was applied to these cells without an additional rule. In this case, $g_i$ becomes positive for $K_i<N_i<A$, and the conventional strong-Allee interpretation does not apply. This low-capacity regime represents a limitation of the present model, as discussed in Sec.~IV.

To maintain nonnegative population states in the numerical implementation, any negative
updated local abundance was set to zero after the population-growth calculation. The same nonnegativity constraint was applied after incorporating movement inflows and outflows. This correction was applied to the updated population state $N_i$, not to the local growth term $g_i$, which may be negative when the population declines.

To model movement between cells of an invasive species, we consider the tendency of individuals to leave the current cell according to density dependence. Individuals are increasingly likely to leave occupied cells as the local population size approaches or exceeds the carrying capacity. We define the number $L_i$ of individuals leaving cell $i$ using the following logistic function:

\begin{align}
L_i&=\frac{N_i}{1+e^{-\beta_\textrm{d}(N_i-K_i)}}.
\end{align}

The number of individuals leaving, $L_i$, increases rapidly as $N_i$ approaches $K_i$, reflecting a greater tendency to emigrate when cell $i$ becomes overcrowded. The parameter $\beta_{\mathrm{d}}$ controls the sensitivity of $L_i$ to changes in $N_i$. A higher value of $\beta_{\mathrm{d}}$ results in a steeper increase in $L_i$ when $N_i$ exceeds $K_i$. In contrast, a lower value of $\beta_{\mathrm{d}}$ produces a smoother change in $L_i$.

Finally, the destination of individuals leaving the current cell $i$ is influenced by habitat preference. A habitat-preference score $p_i$, calculated based on empirical data, is assigned to each cell $i$. We assume that individuals prefer to migrate to cells with higher values of $p_i$ than those of other neighboring cells. We define the difference in habitat score between the destination cell $j$ and the original cell $i$ as $\Delta_{ij}=p_j-p_i$. The movement rate from cell $i$ to cell $j$ is then given by

\begin{align}
	c_{ij} = \frac{1}{\mathcal{Z}} \frac{1}{1+e^{-\beta_\textrm{h}(\Delta_{ij} - \langle{ |\Delta| }\rangle )}},
	\label{mv_rate}
\end{align}

where $Z$ is a normalization factor for $c_{ij}$, defined as
$Z=\sum_{j\in\partial i}\left(1+e^{-\beta_{\mathrm{h}}(\Delta_{ij}-\langle|\Delta|\rangle)}\right)^{-1}$.
The control parameter $\beta_{\mathrm{h}}$ characterizes the sensitivity of movement to differences in habitat preference. Smaller values of $\beta_{\mathrm{h}}$ imply that movement is nearly random, allowing individuals to move to cells with lower habitat scores. Conversely, higher values of $\beta_{\mathrm{h}}$ indicate that individuals are more selective, favoring movement toward cells with higher habitat scores. Thus, $\beta_{\mathrm{h}}$ controls the sensitivity of movement toward more suitable habitats, with higher values leading to more selective migration. Additionally, the midpoint of the logistic function is determined by the mean absolute difference in habitat score between all pairs of neighboring cells. For the study area, the mean absolute difference was calculated as $\langle|\Delta|\rangle=0.17$ using the cell-specific habitat-preference scores derived from the expert-evaluated land-cover scores described in Sec.~II D and Table~\ref{table} in Appendix~A.

\subsection{\textit{Glis glis} (edible dormouse)}

We selected \textit{G. glis}, commonly known as the edible dormouse, as the target species for application of our model. This medium-sized rodent is predominantly found in European countries. It is nocturnal and primarily inhabits trees, showing a preference for broadleaf and mixed forests, as well as orchards and gardens \citep{Koenig1960,	Von1960Siebenschlafer,Michaux2019Genetic,Krystufek2010Glis,Marteau2015habitat}. Given the similarities in climatic conditions within temperate zones and the high rate of trade between Europe and Korea, the edible dormouse may have a relatively high chance of invading Korea. In addition, \textit{G. glis} can maintain access to food throughout seasonal changes by consuming a variety of foods, such as seeds, fruits, and insects, allowing the species to adapt well to non-native and marginal environments \citep{Hurner2009ecology}. Moreover, \textit{G. glis} can act as a host for ticks that transmit Lyme disease \citep{Matuschka1994amplifying,Richter2011reservoir,Fietz2014seasonal}. Owing to its potential impacts on local ecosystems, the edible dormouse was listed among the 200 alert alien species in Korea \citep{book_nie}. For this reason, we apply the dispersal model to \textit{G. glis} to provide a prognosis of its dispersal patterns in the event of invasion into Korea.

\subsection{Study area}

We applied our model to the midwestern region of the Korean Peninsula, spanning latitudes 37.15$^\circ$N--37.62$^\circ$N and longitudes 126.31$^\circ$E--127.26$^\circ$E, covering an area of approximately $84~\mathrm{km}\times53~\mathrm{km}$.

This region was selected because it includes (i) major points of entry for imported animals and goods, (ii) pet-trade-related businesses in urban areas that could serve as potential introduction sources, and (iii) suitable natural habitats surrounding the metropolitan area in which the edible dormouse population could become established. The study area was divided into spatial cells of $180~\mathrm{m}\times180~\mathrm{m}$ ($3.24~\mathrm{ha}$), considering the reported female home-range size of $0.37$--$1.21~\mathrm{ha}$ \citep{Jurczyszyn2007Influence}.

\subsection{Habitat preference score}

We used land-cover data obtained from the Environmental Spatial Information Service \citep{EGIS} in Korea to characterize the habitat types of the spatial cells in the model. These land-cover types were classified into 22 distinct subcategories, as detailed in Table~\ref{table} in Appendix~A. Habitat-preference scores for each land-cover type were obtained through evaluations by field experts conducting mammal surveys in Korea. This expert-based evaluation is consistent with previous studies that empirically determined resistance scores for plant distributions \citep{Liu2018Identifying} and habitat preferences for insect movement \citep{Zhang2023Lattice}. The habitat-preference survey was conducted in two stages, following principles similar to those of the Delphi method \citep{Linstone1975delphi,Macmillan2006Delphi}. In the first stage, each expert independently evaluated the habitat-preference scores. In the second stage, three rounds of reevaluation were conducted, during which experts could revise their assessments after reviewing the mean scores from the preceding round.

\begin{figure}[t]
	\centering
	\includegraphics[width=0.9\columnwidth]{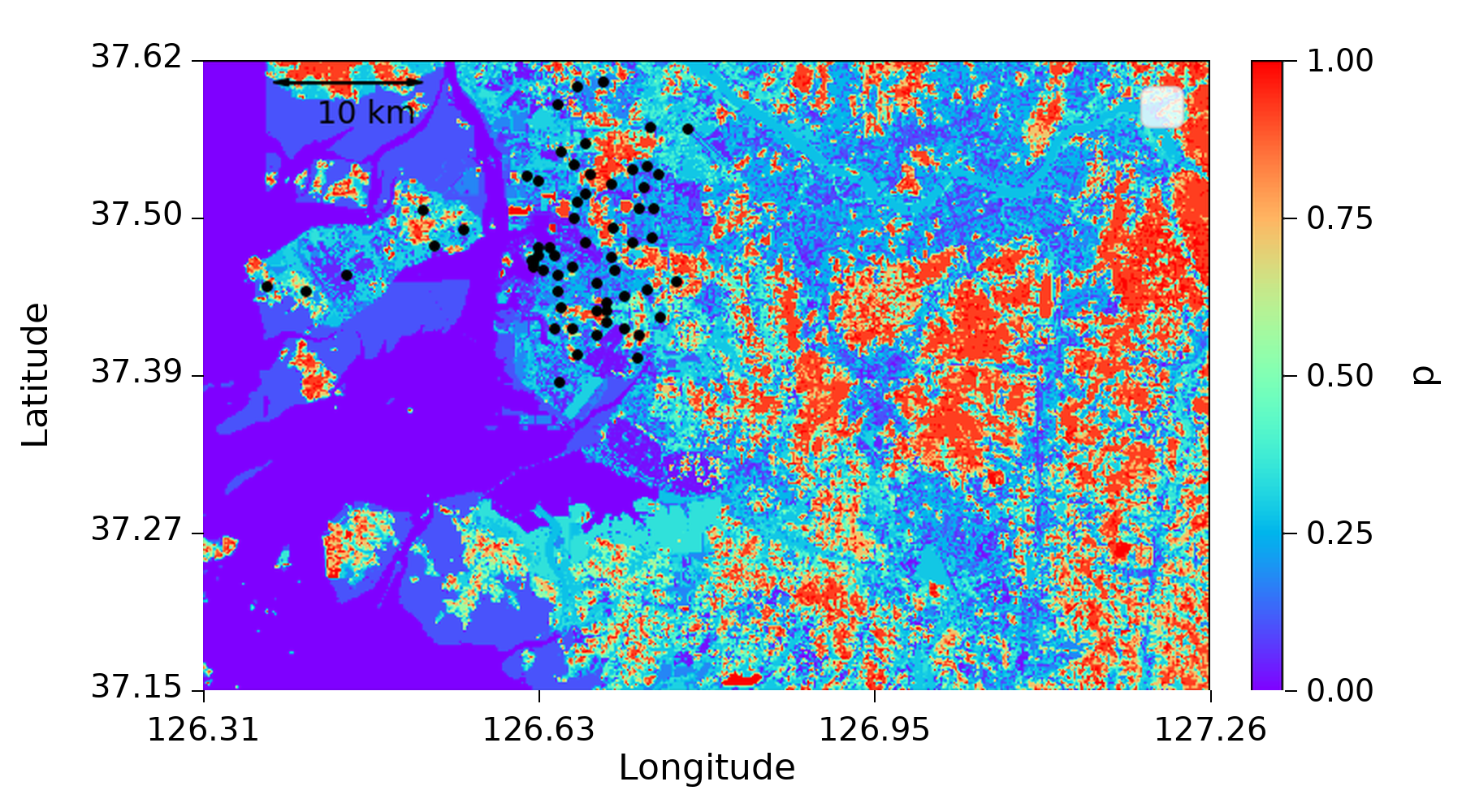}
	\caption{
		Habitat-preference score map of the study area in the midwestern Korean Peninsula. Colors indicate the normalized habitat-preference score $p$, and black dots represent the initial invasion-source locations.
	}
	\label{fig2}
\end{figure}

Table~\ref{table} lists the raw (unnormalized) habitat-preference scores $p_0$ for the 22 distinct land-cover types. We converted these raw scores into normalized habitat-preference scores $p$ using the following equation:
\begin{equation}
	p = \frac{p_0-p_{\mathrm{min}}}{p_{\mathrm{max}}-p_{\mathrm{min}}},
\end{equation}
where $p_{\mathrm{max}}$ and $p_{\mathrm{min}}$ denote the maximum and minimum values of $p_0$, respectively, among the 22 land-cover types. The normalized habitat-preference score $p$ was used in the numerical simulations. The habitat-preference score $p_i$ of each cell was assigned according to its corresponding land-cover type. Figure~\ref{fig2} shows the spatial distribution of the normalized habitat-preference scores over the study area, with relatively high scores occurring broadly in forested areas in the eastern part of the study area.

\subsection{Parameters of Population Dynamics}

Our model includes three parameters governing the dynamical evolution of $N_i$: the potential growth rate $r$, the Allee threshold $A$, and the carrying capacity $K_i$ of each cell. We estimated these parameter values based on empirical data and biologically reasonable assumptions. First, the annual potential growth rate was set to $r=1.72$, as estimated through life-table analysis using an age-class model \citep{Lafever2008,Bieber2018Effects,Bieber2019Repository}. 

Second, we set the Allee threshold to $A=2$ for each cell. According to our literature search, an Allee-effect threshold has not been specifically reported for \textit{G. glis}. This may partly reflect the difficulty of detecting low-density effects under field conditions, particularly when long-term observations are limited. Nevertheless, Allee effects have been documented across a wide range of taxa and ecosystems, including rodents, and may arise through ecological mechanisms such as mate limitation \citep{Courchamp2008Allee,Morris2002Allee,Muir2024Allee}. At least two individuals, one female and one male, are required for sexual reproduction, whereas a single individual or two individuals of the same sex cannot form a mating pair. In this study, we conservatively adopted a low Allee threshold to represent potential establishment failure at very low local abundance. Because sex structure is not explicitly represented in our model, $A=2$ should be interpreted as a phenomenological modeling assumption rather than an empirically estimated threshold for \textit{G. glis} \citep{Boukal2002Allee}. In the present formulation, the local growth term vanishes at $N_i=A$, and the conventional strong-Allee interpretation applies to cells with $K_i>A$. For cells with $0<K_i<A$, the same growth equation is applied without a separate rule, as explained in Sec.~II A.

Finally, we set the maximum carrying-capacity density to $K_{\max}=10$ individuals per hectare based on empirical observations \citep{Jurczyszyn2007Influence}. Because the carrying capacity $K_i$ of cell $i$ is assumed to be proportional to habitat preference, we set $K_i = 3.24\,K_{\max}\,p_i$, where $3.24$ is the cell area in hectares and $p_i$ is the normalized habitat-preference score of cell $i$. Here, $p_i$ is used as a proxy for local habitat quality and resource availability, so $K_i$ represents an effective carrying capacity that varies among cells. The linear proportionality between $K_i$ and $p_i$ is a simplifying modeling assumption rather than an empirically calibrated relationship for \textit{G. glis}. Cells corresponding to sea, for which $p_i=0$ (and thus $K_i=0$), were treated as unsuitable habitat; their population was fixed at $N_i=0$, and movement into these cells was set to zero.

\subsection{Initial invasion sources}

Our main focus in this study is the dispersal of the invasive species described by Eq.~\eqref{rate_eq}. The initial invasion may occur through various pathways, including natural migration and unintentional transport by humans. For the initial stage of edible dormouse dispersal, we consider two major potential sources. 
The first source involves individuals accidentally released during importation through ports, particularly in the western part of the Incheon seaport area. Animals imported through ports as part of the pet trade may serve as potential introduction sources in this scenario. The second source involves imported animals that are subsequently kept as pets. In this case, invasion may result from accidental releases or abandonment by owners, mainly in urban areas, followed by dispersal into nearby forest areas. For the simulations, the potential introduction sources were based on 62 records of animal exhibition and import businesses in the Incheon area, which were mapped onto the model grid.

\subsection{Simulation setup and summary measures}

Based on the initial-invasion scenarios, we performed numerical simulations of Eq.~\eqref{rate_eq}, starting with three individuals at each of the designated source locations shown in Fig.~\ref{fig2}. This initial population was chosen to exceed the Allee threshold of $A=2$. Each simulation covered a period of 100 years with a time step of one year, yielding the population $N_i(t)$ in each cell $i$. 

The annual time step was adopted to maintain consistency with the potential growth rate $r$, which was estimated on a yearly basis from life-table analysis. This temporal resolution is also consistent with the seasonal reproductive cycle of \textit{G. glis}, which may skip reproduction in unfavorable years \cite{Vekhnik2022Reproduction}. Such interannual variation in reproduction is not explicitly represented in the present model. Accordingly, movement between neighboring cells represents effective local population redistribution between annual demographic states, rather than the
total physical distance traveled by an individual during one year. This local-movement approximation is consistent with the high site fidelity reported for adult \textit{G. glis},
although juveniles, particularly males, may disperse over longer distances \cite{Cornils2017Movement}.

We quantified the occupied area and mean population density of the edible dormouse population.
The occupied area $M$ is defined as
\begin{equation}
	M=a\sum_i f(N_i),
\end{equation}
where $a=3.24~\mathrm{ha}$ is the area of one lattice cell, and $f(N_i)$ is the occupancy indicator function defined as
\begin{equation}
	f(N_i)=
	\begin{cases}
		1, & N_i \geq 1, \\
		0, & 0 \leq N_i < 1.
	\end{cases}
\end{equation}
The population size within the occupied area is defined as $N=\sum_i N_i f(N_i)$, and the mean
population density within this area is given by $N/M$ when $M>0$. Thus, both the population sum $N$ and the occupied area $M$ are calculated from the same set of cells with $N_i \geq 1$.

During the deterministic population-growth and movement calculations, however, $N_i$ is retained as a continuous-valued local abundance. Cells with $0<N_i<1$ remain part of the evolving population state but are not included in the reported occupied area or mean population density.

\subsection{Population center and dispersal radius}

To characterize the spatial trajectory and extent of population dispersal, we calculated the
population-weighted center and radius of gyration using the occupancy criterion defined in Eq.~(7). The population center is defined as

\begin{equation}
	\mathbf{x}_{\mathrm{cm}}(t)
	=
	\frac{
		\sum_i \mathbf{x}_i N_i(t) f(N_i(t))
	}{
		\sum_i N_i(t) f(N_i(t))
	},
	\label{eq:center_mass}
\end{equation}

where $\mathbf{x}_i$ is the position vector of cell $i$, and $N_i(t)$ is the population in
that cell at time $t$. The spatial extent of the population around its center is quantified by the radius of gyration,

\begin{equation}
	R(t)
	=
	\sqrt{
		\frac{
			\sum_i N_i(t) f(N_i(t))
			\left\lVert
			\mathbf{x}_i-\mathbf{x}_{\mathrm{cm}}(t)
			\right\rVert^2
		}{
			\sum_i N_i(t) f(N_i(t))
		}
	}.
	\label{eq:gyration}
\end{equation}

Both measures are calculated using only cells with $N_i(t)\geq 1$ and are defined when the
population within the occupied area is positive. The trajectory of $\mathbf{x}_{\mathrm{cm}}(t)$ represents the mean dispersal pathway, whereas $R(t)$ measures the population-weighted spatial spread around that pathway.

\section{Results and Discussion}\label{sec:result}

\subsection{Effect of habitat-preference sensitivity $\beta_{\mathrm{h}}$}

We compared two values of the habitat-preference sensitivity parameter, $\beta_h=2$ and $10$, while fixing the density-dependence sensitivity at $\beta_d=0.31$. The latter value was selected from the middle of the parameter range used in the simulations. We selected $\beta_h=2$ and $10$ to illustrate contrasting degrees of habitat-preference sensitivity, while a broader range of both parameters was explored in Fig.~5. These sensitivity parameters were varied to examine the relative effects of habitat preference and density-dependent movement under the fixed demographic and spatial settings of the model; they were not directly fitted to species-specific dispersal observations.

\begin{figure}[t]
	\centering
	\includegraphics[width=0.85\columnwidth]{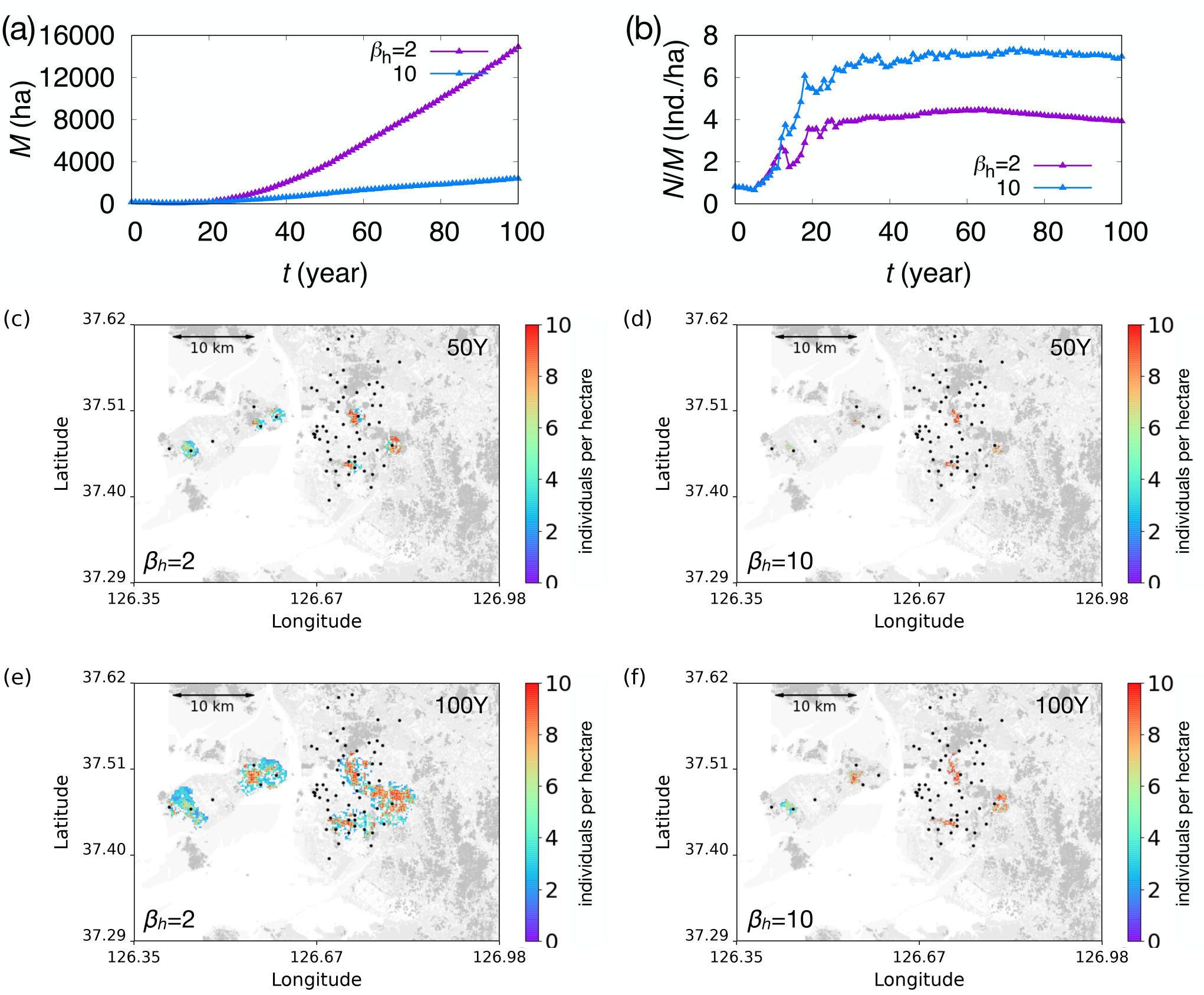}
	\caption{
		Time evolution of (a) the occupied area $M$ and (b) the mean population density $N/M$ for $\beta_{\mathrm{h}}=2$ and $10$, with the density-dependence sensitivity fixed at $\beta_{\mathrm{d}}=0.31$. Panels (c) and (d) show the population-density distributions after 50 years for $\beta_{\mathrm{h}}=2$ and $10$, respectively, whereas panels (e) and (f) show the corresponding distributions after 100 years. The black dots indicate the initial invasion sources, and the colors represent the population density in individuals per hectare. The simulations were performed with $r=1.72$, $A=2$, and $\langle |\Delta| \rangle=0.17$.
	Areas with high population density, shown in red in panels (c)--(f), generally correspond to forested cells with high habitat-suitability scores. The occupied area and mean population density 	are calculated using cells with $N_i\geq 1$, with $N$ representing the population summed 	over these occupied cells.
	}
	\label{fig3}
\end{figure}

Figure~\ref{fig3}(a) shows the occupied area $M$, and Fig.~\ref{fig3}(b) shows the mean population density $N/M$ within the occupied area as functions of time. For both values of $\beta_{\mathrm{h}}$, the occupied area increased throughout the 100-year simulation. The mean population density increased rapidly during the early stage and approached saturation after approximately 30 years.

The two values of $\beta_{\mathrm{h}}$ produced contrasting spatial patterns. Greater habitat-preference sensitivity resulted in a smaller occupied area but a higher mean population density. Under this condition, movement toward cells with higher habitat suitability restricted expansion into less-preferred habitats and concentrated the population in cells with relatively high carrying capacities. In contrast, lower habitat-preference sensitivity allowed the population to occupy a broader range of land-cover types, producing a wider distribution with a lower mean density.

Figures~\ref{fig3}(c)--\ref{fig3}(f) show the simulated population-density distributions after 50 and 100 years. Differences between the two parameter settings were relatively small at year 50 but became more pronounced by year 100. For $\beta_{\mathrm{h}}=2$, the population was distributed over a wider area, whereas for $\beta_{\mathrm{h}}=10$, it remained more spatially localized.

\subsection{Dispersal pathways and radius}

\begin{figure}[t]
	\centering
	\includegraphics[width=0.9\columnwidth]{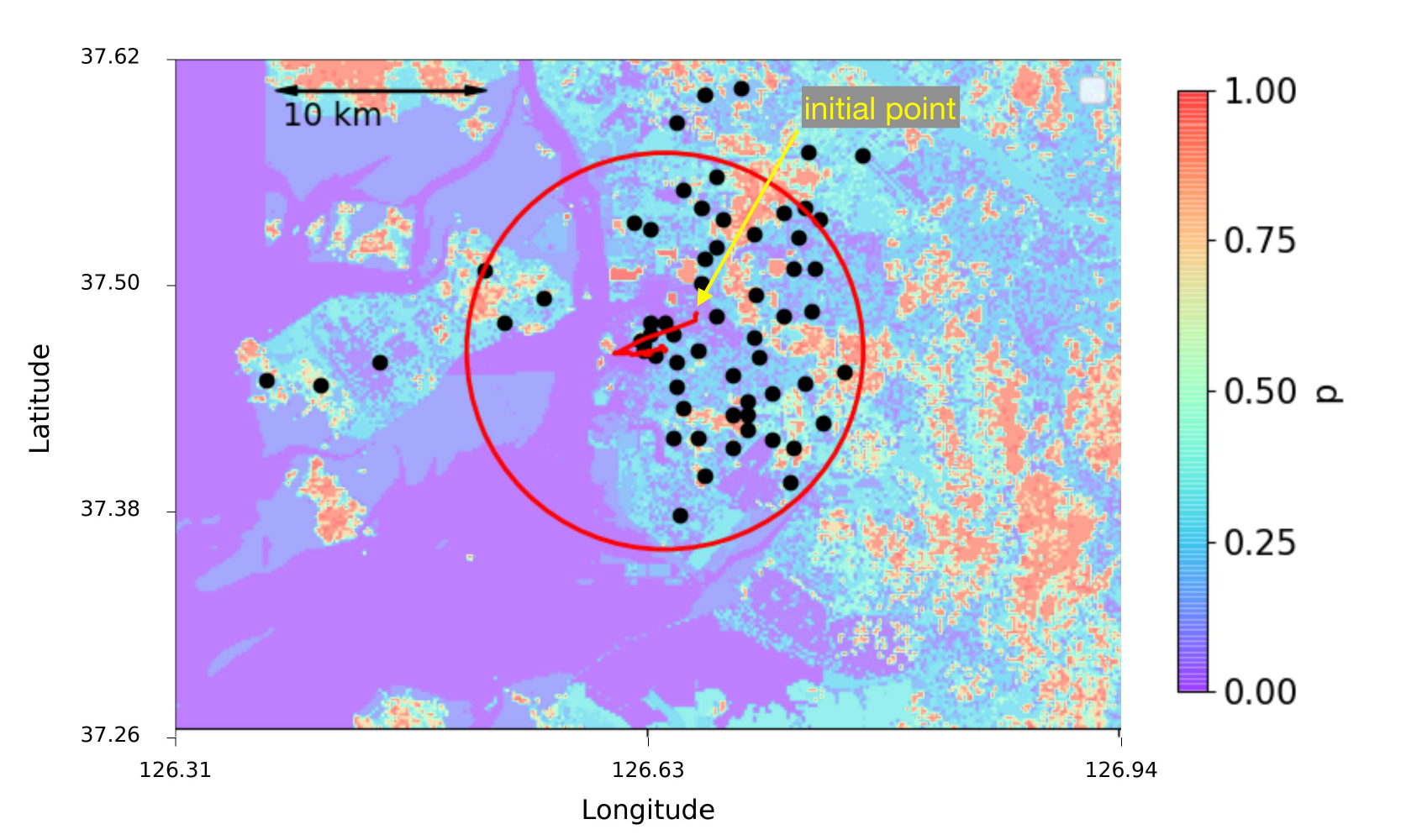}
	\caption{
		The mean dispersal pathway and population-weighted radius of gyration are shown for $\beta_{\mathrm{d}}=0.31$ and $\beta_{\mathrm{h}}=2$, with the initial invasion sources marked by black dots. The red line depicts the trajectory of the population center, and the circle represents the radius of gyration at year 100, rather than the maximum extent of population spread. The parameters used in the simulation were  $r=1.72$, $A=2$, $\beta_{\mathrm{d}}=0.31$, and $\langle|\Delta|\rangle=0.17$. The population center remains close to its initial position throughout the 100-year period. Both the population center and radius of gyration are calculated using cells with $N_i(t)\geq 1$, following Eqs.~\eqref{eq:center_mass} and \eqref{eq:gyration}.
	}
	\label{fig4}
\end{figure}

Using Eqs.~\eqref{eq:center_mass} and \eqref{eq:gyration}, we examined the mean dispersal pathway and spatial extent of the population over the 100-year simulation period (Fig.~\ref{fig4}).

The simulated population center remained close to its initial position, indicating limited net displacement over the 100-year period. This result was somewhat contrary to our expectation that the population center would shift eastward toward the forest-dominated part of the study area shown in Fig.~\ref{fig2}. Although some populations moved eastward during the simulation period, as shown in Figs.~\ref{fig3}(c)--\ref{fig3}(f), populations also persisted in the western part of the study area, causing the population-weighted center to remain near the initial invasion area.

Several features of the simulation may explain this limited shift. First, the local movement scale of the model restricts dispersal to neighboring $180~\mathrm{m}\times180~\mathrm{m}$ cells at annual time steps, resulting in a relatively slow rate of spatial expansion. Second, although forests do not dominate the western part of the study area, this region includes agricultural land and scattered forest cells with relatively high habitat scores, as shown in Table~\ref{table} in Appendix~A. Populations remaining in these western cells offset the contribution of populations that moved eastward when the population center was calculated. And third, the low carrying capacities assigned to urban invasion-source cells may have limited the supply of individuals available to expand toward forested areas in the east during the simulation period. Nevertheless, the population persisted locally around the initial invasion sources in urban areas, indicating that the invaded population could remain viable for at least 100 years under the assumed parameter set.

\subsection{Combined effects of density-dependence and habitat-preference sensitivities}

To examine the combined effects of the two movement mechanisms, we varied the density-dependence sensitivity $\beta_{\mathrm{d}}$ and the habitat-preference sensitivity $\beta_{\mathrm{h}}$. For each parameter combination, we measured the occupied area $M$ and the mean population density $N/M$ after the 100-year simulation period. These quantities characterize the long-term spatial extent and local population level of the established population.

\begin{figure}[t]\centering
\includegraphics[width=\columnwidth]{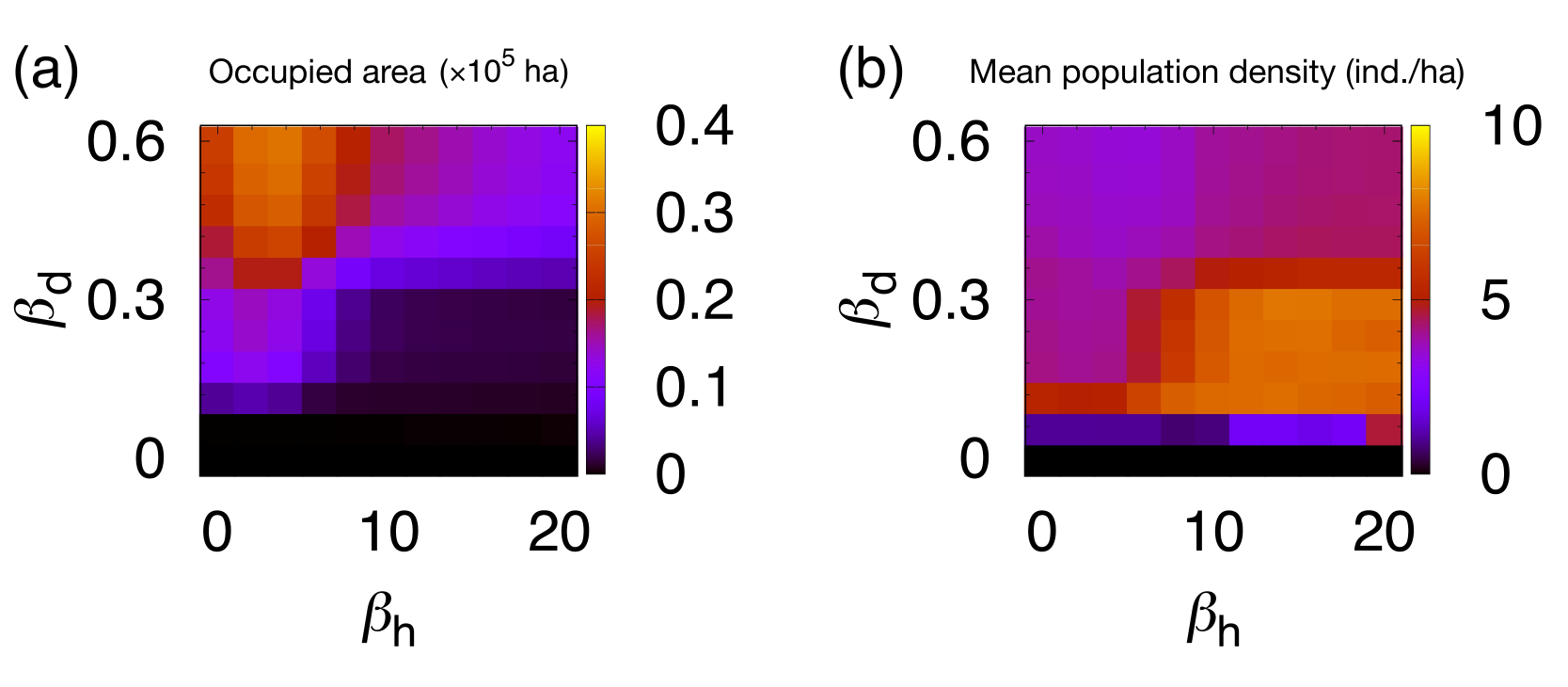}
\caption{
	Occupied area (a) and mean population density (b) after 100 years for various combinations of the habitat-preference sensitivity $\beta_{\mathrm{h}}$ and density-dependence sensitivity $\beta_{\mathrm{d}}$. The simulations used the initial invasion sources shown in Fig.~\ref{fig2}, with $r=1.72$, $A=2$, and $\langle|\Delta|\rangle=0.17$. Occupied area and mean population density are evaluated using cells with $N_i\geq 1$, and the population sum in the mean density includes only these occupied cells. 
}
\label{fig5}
\end{figure}

Figure~\ref{fig5} presents heat maps of the occupied area and mean population density for different combinations of $\beta_{\mathrm{h}}$ and $\beta_{\mathrm{d}}$. The results demonstrate how the interplay between habitat-preference sensitivity and density-dependence sensitivity influences population distribution and growth. In general, greater habitat-preference sensitivity results in a smaller occupied area but a higher mean population density. Conversely, greater density-dependence sensitivity leads to a larger occupied area and a lower mean population density.

The effects of each factor on the occupied area can be analyzed as follows. First, we examine the effect of habitat preference. When $\beta_{\mathrm{h}}$ is low, the population tends to spread over a wide area (Fig.~\ref{fig5}(a)). This is because low habitat-preference sensitivity allows the population to occupy regions with a wide range of habitat scores. In contrast, when $\beta_{\mathrm{h}}$ is high, the population tends to move toward areas with higher habitat scores. This concentration in highly preferred areas causes the population to remain confined to a small number of regions, slowing down the overall spread. As a result, the spatial expansion of the total occupied area is limited.

And second, we consider the effect of density dependence. In our model, movement influenced by density-dependence sensitivity causes local population loss through emigration. This loss is balanced by the inflow of individuals driven by differences in habitat preference. Consequently, the population in occupied cells may increase or decrease depending on the balance between outflow and inflow. When $\beta_{\mathrm{d}}$ is low, the probability of leaving is relatively high even at low population density. As a result, populations can be sustained mainly in regions with high habitat preference, leading to a reduction in the total occupied area (Fig.~\ref{fig5}(a)). In contrast, when $\beta_{\mathrm{d}}$ is high, the probability of leaving becomes relatively high when the population in a cell approaches its carrying capacity. This regulates the rate of outflow and can indirectly promote population persistence and spatial expansion. In this case, populations can survive even in regions with relatively low habitat preference, leading to an increase in the total occupied area.

The mean population density can be examined in relation to the occupied area for different levels of habitat-preference and density-dependence sensitivity (Fig.~\ref{fig5}(b)). In our model, the mean population density is related to the local carrying capacity $K_i$, which is proportional to habitat preference. This means that areas with higher habitat preference are assumed to have larger carrying capacities. As a result, for parameter sets in which movement occurs toward areas with high habitat preference, the mean population density tends to be higher. When habitat-preference sensitivity exceeds 10, the mean population density becomes relatively high. A marked transition in mean population density is observed at $\beta_{\mathrm{d}}=0.31$. Conversely, for parameter sets in which movement allows dispersal into areas with lower habitat preference, the mean population density tends to be lower. Overall, relatively low mean population densities are observed when $\beta_{\mathrm{h}}<10$. In addition, when $\beta_{\mathrm{d}}<0.06$, no significant spatial spread occurs. This suggests that a minimum level of density-dependence sensitivity is necessary for dispersal to occur.

The population dispersal patterns in our numerical results can be categorized into three distinct scenarios based on the parameter values. First, high values of $\beta_{\mathrm{h}}$ and low values of $\beta_{\mathrm{d}}$ result in a localized occupied area with a higher mean population density. Second, low values of $\beta_{\mathrm{h}}$ and high values of $\beta_{\mathrm{d}}$ lead to a widespread occupied area with a lower mean population density. And third, for $\beta_{\mathrm{d}}\leq0.06$, no significant spatial spread occurs. These observations underscore the complex interaction between habitat preference and density dependence in shaping population dispersal.

\section{Conclusion}\label{sec:conclusion}

We demonstrated that population dispersal can be examined within a spatiotemporal framework by incorporating the effects of habitat preference and density dependence into a spatially explicit model. Our model provides insights into dispersal pathways and radii, as illustrated on a geographic map together with measurements of occupied area and population density. From a management perspective, our model provides information on long-term dispersal pathways and radii, which may be useful for developing control measures under specific spatiotemporal conditions. Ecological data are often scarce because field surveys are difficult, and experimental studies are especially constrained for invasive species because deliberate introductions into uninvaded areas are not permissible. Under such conditions, a mechanistic model can provide a transparent, process-based approach for examining potential population dispersal when empirical occurrence data are limited.

Understanding dispersal pathways is crucial for effectively controlling invasive species and clarifying how a species, such as the edible dormouse in this study, expands its occupied area from initial invasion sites to newly established areas. However, it is important to acknowledge that precisely defining the parameter ranges of the logistic functions remains a practical challenge. Long-term observations of the introduced \textit{G. glis} population in Britain provide a useful empirical context. Following its introduction at Tring in 1902, the species was reported to have expanded its distribution by no more than approximately 30 km in any direction by the early 2000s \cite{Morris2004Dormouse}. More recent observations, however, indicate continued local expansion and scattered occurrences at new locations \cite{Trout2021Dormouse}. Although these records document long-term range expansion, they do not provide the habitat-specific movement and local-density information required to estimate
$\beta_h$ and $\beta_d$ independently. Therefore, our simulations are intended to examine relative dispersal patterns across different parameter settings rather than to provide quantitatively calibrated forecasts of geographic spread.

Additionally, realistic values for population growth, such as the net population growth rate, and possible movement distances based on the home range should be investigated in the context of actual dispersal areas. It should be noted that the occupied area was still increasing during the simulation period, although the mean population density reached saturation during the early phase of the simulation. The occupied area may continue to increase over a much longer period, as shown in Fig.~\ref{fig3}(a). The 100-year simulation period was therefore used to examine long-term dispersal scenarios under fixed model assumptions, rather than to predict exact future population distributions. Nevertheless, the nearest-neighbor movement rule does not capture occasional long-distance dispersal, and the annual time step does not explicitly resolve seasonal changes in movement or reproduction.

Carrying capacity has been defined in different ways since its introduction into ecological studies. Traditionally, carrying capacity has often been considered fixed under idealized environmental conditions \cite{Odum1953Ecology}. However, the carrying capacity of a population can vary according to environmental resources and habitat conditions \cite{MonteLuna2004Capacity}. In this study, we assumed that carrying capacity varies according to habitat conditions in each spatial unit. In this way, local density-dependent effects can be more sensitively represented under heterogeneous habitat conditions.

For instance, the maximum carrying-capacity density is $K_{\max}=10$ individuals per hectare, corresponding to 32.4 individuals in a 3.24-ha cell with the maximum normalized habitat-preference score of $p_i=1$. If the habitat-preference score is $p_i=0.5$, the corresponding carrying capacity is 16.2 individuals per cell. For the same local population of $N_i=12$, the ratio $N_i/K_i$ is approximately 0.37 in the former case and 0.74 in the latter, indicating stronger density-dependent limitation in the less suitable habitat. Furthermore, for the same $N_i$ and a fixed positive $\beta_{\mathrm{d}}$, a lower $K_i$ results in a greater emigration fraction $L_i/N_i$ in Eq.~(3), thereby strengthening the density-dependent movement response in less suitable habitats.

Alternatively, carrying capacity could be assumed constant across cells while habitat preference influences only movement between cells. Such an approach is also possible, but it would not explicitly represent differences in local population-growth limits among habitat types. Further empirical data on habitat-specific population abundance and resource availability would be needed to evaluate these alternative modeling assumptions.

A further limitation arises in habitats with very low carrying capacities. Because $K_i$ varies with habitat preference while $A$ is fixed, some cells satisfy $0<K_i<A$. Equation~(2) was applied to these cells without a separate rule, resulting in positive local growth when $K_i<N_i<A$. Consequently, the conventional interpretation of $A$ as a lower Allee threshold and $K_i$ as a stable local carrying capacity does not apply in this regime. Although the numerical implementation prevents negative updated population states, this constraint does not alter the positive-growth interval. The present results should therefore be interpreted as dispersal scenarios under the specified growth formulation. Further analysis would be needed to assess the influence of this low-capacity regime on population density and spatial dispersal.

Additional studies are also required to define the data structure and determine the sizes of spatial and temporal units used in the model. The evaluation of carrying capacity and the Allee-effect threshold in relation to environmental factors under field conditions should also be considered in future studies. Additionally, the model could be applied over a substantially longer period while accounting for climate change, landscape development, and different locations of invasion sources. Nonetheless, our model can help provide a prognosis of the potential extent of spread for species lacking occurrence data, particularly invasive species.

Our model represents population dispersal by describing spatial expansion and temporal growth concurrently. It also links movement mechanisms, including habitat preference and density dependence, to emergent population-level dispersal patterns within a spatially explicit modeling framework.

\section*{ACKNOWLEDGEMENTS}

This research was supported in part by the Korea Environmental Industry \& Technology Institute (KEITI) (No. 2018002270001) through an R\&D project led by the National Institute of Ecology (NIE). This work was also supported by the Global-Learning \& Academic Research Institution for Master's and PhD Students, and Postdocs (LAMP) Program of the National Research Foundation of Korea (NRF), funded by the Ministry of Education (No. RS-2024-00445180).

\clearpage
\appendix
\section{Habitat preference scores}

\begin{table}[!h]
\centering
\small
\renewcommand{\arraystretch}{1.0}
\setlength{\tabcolsep}{5pt}
	
\begin{tabular}{|c|c|c|}

\hline
Land cover type & Sub-land cover type & Score ($p_0$)\\
\hline
\multirow{6}{*}{Urban area}
&Residential area&1.85\\
&Industrial&1.15\\
&Commercial or business area&1.41\\
&Cultural, Sports, and Recreation&1.97\\
&Transportation facilities area&1.24\\
&Public facilities area&1.53\\
\hline
\multirow{5}{*}{Agricultural area}
&Rice Paddy&1.94\\
&Dryland Farming&2.32\\
&Equipped farm land&2.18\\
&Fruit Orchard&2.76\\
&Other cultivated land&2.26\\
\hline
\multirow{3}{*}{Forest area}
&Broadleaf forest&3.82\\
&Coniferous forest&3.21\\
&Mixed forest&4.06\\
\hline
\multirow{2}{*}{Grassland}
&Natural grass&2.53\\
&Artificial grass&2.00\\
\hline
\multirow{2}{*}{Wetland}
&Riparian vegetation&2.12\\
&Coastal wetland&1.41\\
\hline
\multirow{2}{*}{Barren}
&Natural barren&2.18\\
&Artificial barren&1.62\\
\hline
\multirow{2}{*}{Water}
&Inland water&1.91\\
&Sea&1.09\\
\hline
\end{tabular}
\caption{Land cover type and habitat preference scores}
\label{table}
\end{table}

\clearpage
\section*{REFERENCES}

\bibliographystyle{apsrev4-2}
\bibliography{Glisglis_NPSM}

\end{document}